\documentclass{jetpl}
\twocolumn

\title{On a search for the $\eta \rightarrow e^+ e^-$ decay at
the VEPP-2000 $e^+e^-$ collider}
\rtitle{On a search for the $\eta \rightarrow e^+ e^-$ decay at
the VEPP-2000 $e^+e^-$ collider}
\sodtitle{On a search for the $\eta \rightarrow e^+ e^-$ decay at
the VEPP-2000 $e^+e^-$ collider}

\author{M.\,N.\,Achasov$^{+*}$, A.\,Yu.\,Barnyakov$^{+*}$, 
K.\,I.\,Beloborodov$^{+*}$, A.\,V.\,Berdyugin$^{+*}$,A.\,G.\,Bogdanchikov$^{+}$,\\
\vspace{-2mm}
A.\,A.\,Botov$^{+}$, T.\,V.\,Dimova$^{+*}$, 
V.\,P.\,Druzhinin$^{+*}$\/\thanks{e-mail: druzhinin@inp.nsk.su},
V.\,B.\,Golubev$^{+*}$, L.\,V.\,Kardapoltsev$^{+*}$, A.\,G.\,Kharlamov$^{+*}$, 
I.\,A.\,Koop$^{+*\circ}$, L.\,A.\,Korneev$^{+*}$, A.\,A.\,Korol$^{+*}$, 
D.\,P.\,Kovrizhin$^{+*}$, A.\,S.\,Kupich$^{+}$, A.\,P.\,Lysenko$^{+}$,
K.\,A.\,Martin$^{+*}$, 
N.\,Yu.\,Muchnoi$^{+*}$, A.\,E.\,Obrazovsky$^{+}$, E.\,V.\,Pakhtusova$^{+}$,
E.\,A.\,Perevedentsev$^{+*}$, Yu.\,A.\,Rogovsky$^{+*}$, A.\,I.\,Senchenko$^{+*}$,
S.\,I.\,Serednyakov$^{+*}$, Z.\,K.\,Silagadze$^{+*}$, Yu.\,M.\,Shatunov$^{+*}$, 
D.\,A.\,Shtol$^{+}$, D.\,B.\,Shwartz$^{+*}$, A.\,N.\,Skrinsky$^{+}$,
I.\,K.\,Surin$^{+*}$, 
Yu.\,A.\,Tikhonov$^{+*}$, Yu.\,V.\,Usov$^{+*}$, A.\,V.\,Vasiljev$^{+*}$,
and I.\,M.\,Zemlyansky$^{+*}$
}
\rauthor{M.\,N.\,Achasov {\it et al.}}
\sodauthor{M.\,N.\,Achasov {\it et al.}}

\address{\vspace{2mm}$^+$ Budker Institute of Nuclear Physics, SB RAS,
Novosibirsk, 630090, Russia \\
$^*$ Novosibirsk State University, Novosibirsk, 630090, Russia\\
$^\circ$ Novosibirsk State Technical University, Novosibirsk, 630092, Russia}

\abstract{A sensitivity of the VEPP-2000 $e^+e^-$ collider
in a search for the rare decay $\eta \rightarrow e^+ e^-$ has been studied.
The inverse reaction $e^+ e^- \rightarrow \eta$ is proposed for this search.
We have analyzed a data sample with an integrated luminosity of 
108 nb$^{-1}$ collected with the SND detector in the center-of-mass 
energy range $520-580$ MeV and found no background events for the 
reaction $e^+ e^- \rightarrow \eta$ in the decay mode $\eta\to\pi^0\pi^0\pi^0$.
In the absence of background, a sensitivity to 
${\cal B}(\eta \rightarrow e^+ e^-)$ of $10^{-6}$
can be reached during two weeks of VEPP-2000 operation. 
Such a sensitivity is better than the current upper limit on
${\cal B}(\eta \rightarrow e^+ e^-)$ by a factor of 2.3.}

\begin{document}
\maketitle

\begin{figure}
\centering
\includegraphics[width=70mm]{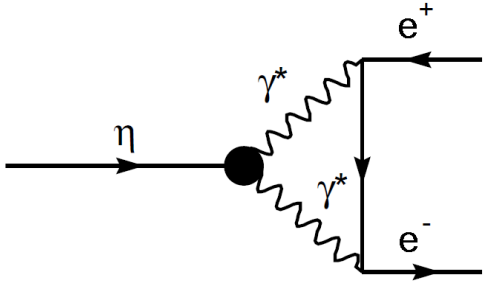}
\caption{{\rm Figure 1}. The Feynman diagram for $\eta\to e^+e^-$.
\label{diag}}
\end{figure}

%We perform a feasibility study  of a search for the  
%$\eta\to e^+e^-$ decay at the $e^+e^-$ VEPP-2000 collider (BINP, Novosibirsk). 
Decays of pseudoscalar mesons to the pair of leptons $P\to l^+l^-$ are rare. 
In the Standard Model (SM) these decays proceed through the two-photon 
intermediate state as shown in Fig.~\ref{diag} and therefore are 
suppressed by a factor of $\sim \alpha^2$ comparing with the
$P\to \gamma\gamma$ decays, where $\alpha$ is the 
fine structure constant.
An additional suppression of $(m_l/m_P)^2$ arises from the approximate
helicity conservation, where $m_l$ and $m_P$ are the lepton and meson masses,
respectively. So, the $P\to l^+l^-$ partial width is less 
than the two-photon width $\Gamma(P\to \gamma\gamma)$ by a factor of 
$\sim \alpha^2(m_l/m_P)^2$.
The low probability makes these decays sensitive to possible contributions 
of new physics beyond the SM~\cite{rare1,rare2}.
In the SM, the knowledge of the transition meson-photon
form factor $F(q_1^2,q_2^2)$ for the $\gamma^\ast\gamma^\ast\to P$
vertex is needed for a $\Gamma(P\to l^+ l^-)$ calculation,
where $q_1^2$ and $q_2^2$ are four-momenta squared of virtual 
photons in Fig.~\ref{diag}. The real and imaginary parts of 
the decay amplitude are usually calculated separately. The imaginary part 
$\Im(A)^2$ is proportional to the form-factor value $|F(0,0)|^2$
at $q_1^2=q_2^2=0$ and, consequently, can be calculated using the known 
width of the $P \rightarrow \gamma \gamma$ decay. Since $|A|^2 > \Im(A)^2$, 
a model-independent lower boundary (unitary bound) on 
the $P \rightarrow l^+ l^-$ width can be obtained from 
$\Gamma(P\to \gamma\gamma)$~\cite{ul}.
The real part of amplitude cannot be calculated in a model-independent
way. In Ref.~\cite{ReA} it is shown that the real part depends
on the integral over $q^2<0$ of the form factor in symmetric kinematics, 
$F(q^2,q^2)$. The ranges of predictions for the 
$P \rightarrow l^+ l^-$ branching fractions obtained in different 
form-factors models~\cite{th1,th2} are listed in Table~I. 
For comparison, the last column of Table~I contains the 
current experimental values of the branching fractions.
\begin{table*}
\begin{center}
\caption{\label{tab1}{\rm Table I.} 
The unitary bounds, 
theoretical predictions in units of the unitary bound~\cite{th1,th2},
and experimental values for the $P \rightarrow l^+ l^-$ branching fractions. 
}
\begin{tabular}{cccc}
\hline
${\cal B}(P \rightarrow l^+ l^-)$ & Unitary bound ($U$) &  Theory & Experiment \\
\hline
${\cal B}(\pi^0 \rightarrow e^+ e^-) \times 10^8$             &$ 4.69$ &$6.23-6.38$  & $7.49 \pm 0.38$~\cite{KTeV}    \\
${\cal B}(\eta \rightarrow e^+ e^-)\times 10^9$               &$ 1.78$ &$4.60-5.24$  & $<2300$~\cite{hades}    \\
${\cal B}(\eta \rightarrow \mu^+ \mu^-)\times 10^7$           &$ 4.36$ &$4.64-5.12$  & $5.8 \pm 0.8$ ~\cite{pdg}    \\
${\cal B}(\eta^{\prime} \rightarrow e^+ e^-)\times 10^{10}$   &$ 0.36$ &$1.15-1.86$  & $< 56$ ~\cite{snd,cmd3} \\
${\cal B}(\eta^{\prime} \rightarrow \mu^+ \mu^-)\times 10^7$  &$ 1.35$ &$1.14-1.36$  & $-$   \\
\hline
\end{tabular}
\end{center}
\end{table*}
Currently, only two of the five decays are measured and for two more 
upper limits on the width exist. 
The value of ${\cal B}(\eta \rightarrow \mu^+ \mu^-)$ agrees with the 
prediction but the measurement 
accuracy is low. The more precise value of
${\cal B}(\pi^0 \rightarrow e^+ e^-)$  differs from the prediction
by about three standard deviations.

It is clear that the current experimental situation requires improvements
in accuracy for the $\pi^0 \rightarrow e^+ e^-$ and 
$\eta \rightarrow \mu^+ \mu^-$ decays 
as well as measurements of the other three decays. Such measurements  
are planned at the BES-III~\cite{bes}, KLOE-2~\cite{kloe}, 
Crystal Ball~\cite{cb} and WASA~\cite{wasa1,wasa2} 
detectors in the near future.

The upper limit on ${\cal B}(\eta^{\prime} \rightarrow e^+ e^-)$ 
has been recently set 
in experiments with the CMD-3~\cite{cmd3} and SND~\cite{snd} at VEPP-2000,
where the inverse reaction $e^+ e^- \to \eta^{\prime}$ was used. In this paper 
we propose the same method for a search for the $\eta \rightarrow e^+ e^-$ 
decay and perform its feasibility study.

%\section{The VEPP-2000 $e^+e^-$ collider and SND detector}
Data used in this paper were collected with the SND detector
at the VEPP-2000~\cite{vepp} collider in 2013.
VEPP-2000 is designed for a study of $e^+e^-$ annihilation at the 
center-of-mass (c.m.) energy ($E$) from 160 MeV up to 2 GeV. There are two 
detectors
at the collider, SND and CMD-3, which collect data simultaneously. 
At the moment of data taking the VEPP-2000 accelerator complex consisted
of the 3 MeV electron linac ILU, the 250 MeV pulsed synchrotron B-3M,
the 900 MeV booster storage ring BEP, and the collider storage ring VEPP-2000.

Currently the complex VEPP-2000 is being upgraded. The maximum energy of  
BEP will be increased up to 1 GeV, and the injection system will be changed. 
Electrons and positrons will be transported to BEP from the VEPP-5 
injection complex~\cite{vepp5} through a 250 m beamline.
The upgrade allows to increase the VEPP-2000 luminosity
at maximum energy up to $10^{32}$ cm$^{-2}$sec$^{-1}$ and should result in a more 
stable operation of the accelerator complex.

For the current study, VEPP-2000 parameters at c.m. energy 
close to $m_\eta c^2=548.862\pm0.018$ MeV~\cite{pdg} such as 
luminosity, accuracy of the energy setting, energy spread, 
are important.
In 2013 SND did not record data exactly at this energy.
Therefore, we analyze data from four energy points
near $m_\eta c^2$, with c.m. energies of 520, 540, 560, and 580 MeV. 
The integrated luminosity collected at these energy points is measured using
the reaction $e^+e^-\to \gamma\gamma$ to be $ 108.1\pm2.0$ nb$^{-1} $.
The average luminosity during data taking varied from 
$0.26\times 10^{30}$ cm$^{-2}$sec$^{-1}$ at 520 MeV to
$0.73\times 10^{30}$ cm$^{-2}$sec$^{-1}$ at 580 MeV.
Interpolating the energy dependence we estimate that the 
average luminosity expected at $E=m_\eta c^2$ is 
$0.34\times 10^{30}$ cm$^{-2}$sec$^{-1}$.
% From the distribution of the integrated luminosity 
%over the energy points and effective time of data collection
%we estimate that average luminosity
%expected at $E=m_\eta c^2$  is $0.34\times 10^{30}$ cm$^{-2}$sec$^{-1}$.
 
The width of the $\eta$ resonance, $\Gamma_\eta=1.31\pm0.05$ keV~\cite{pdg},
is much less than the collider c.m. energy spread $\sigma_E$.
This means that the value of $\sigma_E$ is crucial for the proposed 
search for the $e^+e^-\to\eta$ reaction. At 
$E=m_\eta c^2$ the energy spread is determined by the Touschek effect. 
It is uniquely related to the RMS of the longitudinal distribution 
of the $e^+e^-$ interaction point $\sigma_{Z}$~\cite{cmd3}:
\begin{equation}
\sigma_{E} = 4.05\sigma_{Z}\sqrt{V_{\rm cav}E_{\rm b}
\sin[\arccos(63.2E_{\rm b}^4/V_{\rm cav})]},
\label{eq4}
\end{equation}
where $\sigma_{E}$ is measured in keV, $\sigma_{Z}$ in mm,
the RF cavity voltage $V_{\rm cav}$ in kV, and the
beam energy $E_{\rm b}=E/2$ in GeV.
The value of $\sigma_{Z}$ is measured using detected events of the elastic
scattering  $e^+e^-\to e^+e^-$.
At $E=540$ MeV data taking was carried out with several values of 
the RF cavity voltage ranged from 18 kV to 36 kV. The values
of $\sigma_{Z}$ varied from 16 to 12 mm, while the values of $\sigma_{E}$
from 120 to 160 keV. Below we will use the value of $ \sigma_E = 150$ keV
as an estimate of the energy spread at $E=m_\eta c^2$.

In the proposed experiment the collider energy must be set 
and monitored with an accuracy better than $\sigma_{E}$.  
This is provided by the beam-energy-measurement system based on 
measuring energies of Compton back-scattered laser photons~\cite{emes}.
The accuracy of this method is about 60 keV. The system monitors beam energy
continuously during data taking; the single measurement duration at 
$E\approx 550$ MeV is about one hour.

The search for $e^+ e^- \to \eta$ events will be performed with the
SND detector~\cite{det1,det2,det3,det4}. It is a nonmagnetic detector. 
The main part of the detector is a three-layer spherical
electromagnetic calorimeter consisting of 1640 NaI(Tl) crystals.
The calorimeter covers a solid angle of about 90\% of 4$\pi$.
The energy and angular resolutions for photons with energy $E_\gamma$ 
are described by the following formulae 
\begin{eqnarray}
\sigma_{E_\gamma}/E_\gamma&=&4.2\%/\sqrt[4]{E_\gamma(\mbox{GeV})},\\
\sigma_{\theta,\phi}&=&0.82^{\circ}/\sqrt{E(\mbox{GeV})}.
\end{eqnarray}
Directions of charged particles are measured in a nine-layer drift chamber.
The calorimeter is surrounded by an iron absorber and a muon detector. 
In the proposed search the muon detector is used to veto cosmic rays. 

%\section{The cross section for $e^+e^-\to \eta$}
The energy dependence of the Born cross section for the reaction 
$e^+ e^- \rightarrow  \eta$ is described by the Breit-Wigner formula:
\begin{equation}
\sigma_{0} =\frac{4\pi}{E^2} {\cal B}(\eta \rightarrow e^+ e^-)
\frac{m^2_{\eta}\Gamma^2_{\eta }}{(m_{\eta}^2-E^2)^2+m_{\eta}^2\Gamma_\eta^2}.
\label{eq1}
 \end{equation}
In experiment, it is necessary to take into account radiative 
corrections arising, for example, from radiation of extra photons from the 
initial state. To do this, the cross section (\ref{eq1}) is convolved
with the so-called radiator function $W(s,x)$~\cite{rad1,rad2} 
\begin{equation}
\sigma(s)= \int_{0}^{x_{max}}W(x,s)\sigma_{0}(s(1-x))dx, \label{eq2}
\end{equation}
where $s=E^2$. The upper limit of integration in Eq.(\ref{eq2}) depends on
the decay mode and equals unity for the $\eta$ decay into two photons and 
$1-{(3m_{\pi^0})^2}/s$ for the decay into $3\pi^0$. The theoretical accuracy 
of the corrected cross section (\ref{eq2}) is better than 1\%~\cite{rad1,rad2}.
For the unitary bound ${\cal B}(\eta \rightarrow e^+ e^-)=1.78\times 10^{-9}$
the Born cross section in the resonance maximum is $\sigma_0=29$ pb. 
The radiative corrections lead to decrease of this cross section 
down to $\sigma=14$ pb.

To take into account the collider energy spread, the 
cross section (\ref{eq2}) should be convolved with the Gaussian function
describing the distribution of the integrated luminosity over energy
\begin{equation}
 \sigma_{\rm exp}(E_{0}) = \frac{1}{\sqrt{2\pi}\sigma_E}
 \int\limits_{-\infty}^{+\infty}e^{-\frac{(E-E_{0})^2}{2\sigma^2_E}}\sigma(E) dE,
 \label{eq5}
 \end{equation}
where $E_{0}$ is the average beam energy. 
For $\sigma_E = 150$ keV, $E_0=m_\eta c^2$, and
${\cal B}(\eta \rightarrow e^+ e^-)=1.78\times 10^{-9}$
the visible cross section is 
\begin{equation}
\sigma_{\rm exp}^U(m_\eta c^2)=105\pm 11\mbox{ fb}.
\label{xs-ul}
\end{equation}
Thus, the energy spread allows to use only 
$\sigma_{\rm exp}/\sigma\approx 1/140 $ of collected integrated 
luminosity. 
The uncertainty on $\sigma_{\rm exp}^U(m_\eta c^2)$ is due to
the uncertainty of the beam-energy measurement (60 keV).
 
%\section{Estimation of background for $e^+e^-\to \eta$}
The main $\eta$ decay modes are
$ \eta \rightarrow \gamma  \gamma $ (39.4\%),
$ \eta \rightarrow \pi^0  \pi^0  \pi^0 $ (32.7\%) and
$ \eta \rightarrow \pi^+  \pi^-  \pi^0 $ (22.9 \%). The numbers in 
parentheses are the branching fractions~\cite{pdg}.
In the $ \eta \rightarrow \gamma  \gamma $ mode 
the process $e^+e^-\to \eta$ cannot be separated from the QED process
$e^+e^-\to \gamma  \gamma $.

The most suitable $\eta$ decay mode for the search for the $e^+e^-\to \eta$
reaction at SND is $\eta \rightarrow \pi^0  \pi^0  \pi^0 \to 6\gamma$,
for which physical background is small.  The main source of background is 
cosmic-ray events. 
For the search for $e^+e^-\to \eta$, events with six or more 
detected photons and with the energy deposition in the calorimeter larger
than $0.6 E$ are selected.
Background from events with charged particles is rejected by the 
selection condition
that the number of fired wires in the drift chamber is less than four. 
Cosmic-ray background is suppressed by the veto from the muon detector.

\begin{figure}
\centering
\includegraphics[width=70mm]{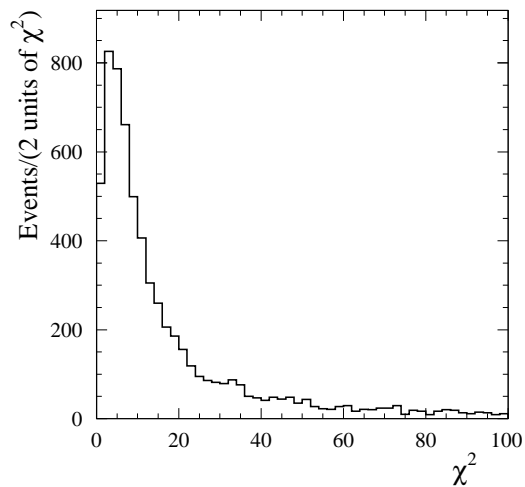}
\caption{{\rm Figure 2}. The distribution of $\chi^2$ of the 
kinematic fit for simulated $e^+e^- \rightarrow \eta$ events 
passing the selection criteria for six-photon events.
\label{fig2}}
\end{figure}

For events passing the preliminary selection, a kinematic fit to the
$e^+e^- \rightarrow \pi^0  \pi^0  \pi^0 \rightarrow 6\gamma $
hypothesis is performed with a requirement of total energy and momentum
conservations and a condition that invariant masses of three photon pairs 
are equal to $\pi^0$ mass. The quality of the kinematic fit is characterized
by the parameter $\chi^2$. During the fit all possible combinations 
of photons are checked and a combination with the smallest $\chi^2$ value 
is selected. The $\chi^2$ distribution for simulated events of the process
$e^+e^- \rightarrow \eta$ passed the selection criteria for six-photon events
is shown in Fig~\ref{fig2}. The condition $\chi^2<100$ is used to
select $\eta$ candidates.

The detection efficiency for $e^+e^- \rightarrow \eta$ events 
is obtained using simulation to be 
$\varepsilon=(12.5\pm 0.6)\%$. The quoted error is systematic. 
It is estimated using results of Ref.~\cite{Achasov:2013btb},
where comparison of data and simulated $\chi^2$ distributions
was performed for five-photon events from the process 
$e^+e^- \to \omega\pi^0 \to \pi^0\pi^0\gamma$.
 
No events satisfying the six-photon selection criteria are
found in the data sample with an integrated luminosity of
$ 108.1\pm2.0$ nb$^{-1}$ recorded
in the energy points with $E=520$, 540, 560 and 580 MeV.

Main sources of physical background for the decay mode
$\eta \to \pi^0  \pi^0 \pi^0$ are five-photon events of
the processes $e^+ e^- \rightarrow \pi^0  \pi^0 \gamma$ and 
$e^+e^-\to 5 \gamma$ with a spurious photon from 
beam-generated background or splitting
of electromagnetic showers in the calorimeter.
The probability to find a spurious photon in an event
may reach up to 5\%. This number is used below for background estimation. 
The cross section for $e^+ e^- \to \pi^0  \pi^0 \gamma $
at $E\approx 550$ MeV is determined mainly by the transition
$e^+ e^- \to \rho(770) \to\pi^0  \pi^0 \gamma $ and is estimated
to be $2\pm1$ pb. The quoted uncertainty is due to an unknown
mechanism of the the $\rho\to\pi^0  \pi^0 \gamma$ decay~\cite{ppg}.
The detection efficiency for $e^+ e^- \rightarrow \pi^0  \pi^0 \gamma $ 
events with the six-photon selection criteria is about 1\%.
The cross section for $e^+e^-\to 5 \gamma$ with detection
of all five photons is calculated using the CompHep program~\cite{comphep}
to be about 4 pb. The requirement of an extra photon and condition 
$\chi^2<100$ decrease the cross section by a factor of about 1/200.
Thus, the total cross section of background processes with
the selection criteria for the decay mode 
$\eta \to \pi^0  \pi^0 \pi^0$ is about 0.04 pb and
negligible at integrated luminosities, which can be collected
at VEPP-2000. 

The second decay mode suitable to search for $ e^+  e^-  \rightarrow \eta$
events is $ \eta \rightarrow \pi^+  \pi^-  \pi^0 $. For this mode
there is a background from the nonresonant process
$e^+  e^-  \rightarrow \pi^+  \pi^-  \pi^0 $.
There are no data on the $e^+  e^-  \rightarrow \pi^+  \pi^-  \pi^0 $
cross section at $E\approx 550$ MeV. To estimate it,
we use the results of Ref.~\cite{snd3pi}, where the cross section
$ e^+  e^-  \rightarrow \pi^+  \pi^-  \pi^0 $ was measured 
by the SND detector at the VEPP-2M collider in the energy region
of the $\omega(782)$ and $\rho(770)$ resonances. 
In Ref.~\cite{snd3pi} the data were fitted in
several models, which at $E=550$ MeV give the cross section value
ranged from 6 to 16 pb. Such cross section values  correspond
to branching fractions 
${\cal B}(\eta \rightarrow e^+ e^-)=(0.4-1.2)\times 10^{-6}$.
Thus, at the level of sensitivity to ${\cal B}(\eta \rightarrow e^+ e^-)$
of $10^{-6}$ the measurement of the $e^+ e^-  \to \pi^+  \pi^-  \pi^0 $
out of the $\eta$-meson resonance may be required.
 Other background sources in the mode $\eta\rightarrow \pi^+  \pi^-  \pi^0 $
are the QED processes 
$ e^+  e^-  \rightarrow e^+  e^- \gamma ( \gamma )$ and
 $ e^+  e^-  \rightarrow \mu^+  \mu^- \gamma ( \gamma)$.
In a data sample collected with the SND detector at $E=520$, 540, 560,
and 580 MeV, we cannot suppress background from these processes
to the level reached in the mode $e^+e^- \rightarrow \pi^0  \pi^0 \pi^0 $.

% \section{\bf\boldmath Оценка чувствительности эксперимента на ВЭПП-2000
% для поиска распада $\eta \rightarrow e^+ e^-$} сечение процесса $ e^+ e^-\rightarrow  \eta $ вычисляется как
The measured cross section $e^+e^- \rightarrow \eta$ is determined as
\begin{equation}
\sigma_{\rm exp} = \frac{N_s}{\varepsilon L},
\label{eq7}
\end{equation}
where $N_s$ is the number of selected $ e^+ e^-\rightarrow  \eta $ events,
$\varepsilon$ is the detection efficiency, and $L$ is the integrated 
luminosity.
 
From our result that no background events for the decay mode 
 $\eta \rightarrow \pi^0  \pi^0 \pi^0 $ are found in the data sample with 
an integrated luminosity of 108.1 nb$^{-1}$ we estimate the
sensitivity of the $ e^+ e^-\rightarrow  \eta $ cross 
section measurement. The upper limit on the cross section
at the 90\% confidence level (CL)~\cite{CoHi} 
corresponding to $N_s=0$ is
\begin{equation}
\sigma_{\rm exp} < \frac{2.3}{0.125 \cdot 108} = 170\mbox{ pb}.
\label{eq8}
\end{equation}

Comparing this limit with the cross section (\ref{xs-ul}) calculated for
the unitary bound ${\cal B}(\eta \rightarrow e^+ e^-)=1.78\times 10^{-9}$,
we estimate the sensitivity to the search for the decay 
$\eta \rightarrow e^+ e^-$ with an integrated luminosity of 108 nb$^{-1}$ 
to be
\begin{equation}
{\cal B}(\eta \rightarrow e^+ e^-) < 2.9 \times 10^{-6}
\label{eq9}
\end{equation}
at 90\% CL. This result is close to the upper limit 
${\cal B}(\eta \rightarrow e^+ e^-)<2.3 \times 10^{-6}$
set recently in the HADES experiment~\cite{hades}, in which
$\eta$ mesons were produced in proton-nucleon collisions.
With a VEPP-2000 luminosity of $0.34\times10^{30}$cm$^{-2}$sec$^{-1}$ the 
current upper limit can be reached in a week of data taking. In two 
weeks a sensitivity at the level of $10^{-6}$ can be reached.

In conclusion, we have analyzed the data sample with an integrated luminosity
of 108 nb$^{-1}$ collected with the SND detector at the VEPP-2000 $e^+e^-$ 
collider at c.m. energies 520, 540, 560, and 580 MeV and found no 
background for the process $e^+ e^- \rightarrow \eta$ in the decay mode 
$\eta\to\pi^0\pi^0\pi^0$.
In the absence of background, data with an integrated luminosity of 324 
nb$^{-1}$ provide a  sensitivity of $10^{-6}$ for 
${\cal B}(\eta \rightarrow e^+ e^-)$.
Such data may be accumulated in two 
weeks of VEPP-2000 operation. The sensitivity of $ 10^{-6}$ is
2.3 times better than the current upper limit on
${\cal B} (\eta \rightarrow e^+ e^-)$.

Data at $E=m_\eta c^2$ will be recorded during the experiment 
on a measurement of hadronic cross sections below 1 GeV planned at VEPP-2000. 
Two detectors,
SND and CMD-3, will collect data simultaneously. Their results 
may be combined to improve sensitivity
to ${\cal B}(\eta \rightarrow e^+ e^-)$ by a factor of about 2. 

We thank S.I. Eidelman for useful discussions. Part of this
work related to the photon reconstruction algorithm in the
electromagnetic calorimeter for multiphoton events is supported
by the Russian Science Foundation (project No. 14-50-00080).
This work is partly supported by the RFBR grant No. 15-02-03391.

\end{document}